\documentclass[twocolumn,floatfix,prl,showpacs,preprintnumbers,footinbib]{revtex4}
\usepackage[dvips]{graphicx}
\usepackage{dcolumn}
\usepackage{bm}
\usepackage[usenames]{color}

\begin{document}

\title{On the validity of the Franck-Condon principle in the optical
spectroscopy: optical conductivity of the Fr\"{o}hlich polaron}
\author{G. De Filippis$^1$, V. Cataudella$^1$, A. S. Mishchenko$^{2,3}$, C.
A. Perroni$^4$, and J. T. Devreese$^5$}
\affiliation{$^1$ Coherentia-INFM and Dip. di Scienze Fisiche - Universit\`{a} di Napoli
Federico II - I-80126 Napoli, Italy\\
$^2$CREST, Japan Science and Technology Agency (JST) - AIST - 1-1-1 -
Higashi - Tsukuba 305-8562 - Japan\\
$^3$RRC ``Kurchatov Institute'' - 123182 - Moscow - Russia\\
$^4$Institute of Solid State Research (IFF), Research Center J\"{u}lich, J\"{u}lich,
D-52425, Germany\\
$^5$TFVS, Departement Fysica - Universiteit Antwerpen - Universiteitsplein 1
- B2610 Antwerpen - Belgium}
\date{\today}

\begin{abstract}
The optical absorption of the Fr\"{o}hlich polaron
model is obtained by an approximation-free Diagrammatic Monte Carlo method and
compared with two new approximate approaches that treat lattice relaxation effects
in different ways. We show that: i) a
strong coupling expansion, based on the the Franck-Condon principle, well
describes the optical conductivity for large coupling strengths ($\alpha >10$);
ii) a Memory Function Formalism with phonon broadened levels reproduces the optical
response for weak coupling strengths ($\alpha <6$) taking the dynamic lattice relaxation into
account. In the coupling regime $6<\alpha<10$ the optical conductivity is a rapidly
changing superposition of both Franck-Condon and dynamic contributions.
\end{abstract}

\maketitle

The Franck-Condon (FC) principle\cite{1dev} and its validity have been
widely discussed in studies of optical transitions in atoms, molecules\cite%
{2dev} and solids\cite{3dev}. For instance, studies on the dynamics
governing color center spectra, Mossbauer spectra\cite{4dev}, and tunneling
in polarizable media\cite{5dev} have established the limits of applicability
of {\ the} FC principle in solids. Generally, {\ the} FC principle means
that if only one of two coupled subsystems, e.g. {\ an} electronic {\
subsystem}, is affected by {\ an} external perturbation, the second
subsystem, e.g. {\ the} lattice, is not fast enough to follow {\ the}
reconstruction of {\ the} electronic configuration. In the opposite limit,
when the perturbation is slow or even static, the characteristic time of
lattice interconfigurational coupling $\tau _{ic}$ is short enough for {\ the%
} lattice to follow slow {\ changes} of {\ the} electronic state
dynamically. It is clear that the justification for {\ the} FC principle is
the short characteristic time of {\ the} measurement process $\tau _{mp}\ll
\tau _{ic}$, where $\tau _{mp}$ is related to the energy gap between the
initial and final states, $\Delta E$, through {\ the} uncertainty principle: 
$\tau _{mp}\simeq \hbar /(\Delta E)$. Then, the spectroscopic response
considerably depends on the value of the ratio $\tau _{mp}/\tau _{ic}$. For
example, in mixed valence systems, where {\ the} ionic valence fluctuates
between {\ the} configurations $f^{5}$ and $f^{6}$ with characteristic time $%
\tau _{ic}\approx 10^{-13}$s, the spectra of fast and slow experiments
are dramatically different\cite{Khomsky,Falicov}. Photoemission experiments
with short characteristic times $\tau _{mp}\approx 10^{-16}$s (FC regime),
reveal two lines, corresponding to $f^{5}$ and $f^{6}$ states. On the other
hand slow M\"{o}ssbauer isomer shift measurements with $\tau
_{mp}\approx 10^{-9}$s show a single broad peak with mean frequency
between signals from pure $f^{5}$ and $f^{6}$ shells. Finally, according to 
the paradigm of measurement process time, magnetic neutron
scattering with $\tau _{mp}\approx \tau _{ic}$ revealed both coherent lines,
with all subsystems dynamically adjusted{,} and broad incoherent remnants
of strongly damped excitation of $f^{5}$ and $f^{6}$ shells\cite{AleksMe95}.
Actually, the meaning and the definition of the times $\tau _{ic}
$ and $\tau _{mp}$ varies with the system and with the measurement
process although the spectroscopic response essentially depends on
whether one of the two interacting subsystems is adjusted to the
changes, induced by a measurement probe, in the other subsystem.

To investigate the interplay between the measurement process time $%
\tau_{mp}$ and the adjustment time $\tau_{ic}$, we study in this Letter
the optical conductivity (OC) of a paradigmatic model for electron-phonon (e-ph)
interaction: the Fr\"ohlich model. Our aim is to investigate the OC from the weak to the
strong coupling regime (in this model the e-ph coupling strength is controlled
by the dimensioless parameter $\alpha$) by three methods: i) the diagrammatic Monte Carlo
(DMC) method \cite{Andrei,Andrei1} which gives numerically exact answers
in all e-ph coupling regimes; ii) the memory
function formalism (MFF) which is able to take dynamical lattice relaxation
into account; iii) and a strong coupling expansion (SCE) which assumes
the FC principle. In the coupling regime $6< \alpha <10$ we find that the
OC spectrum exhibits two features with different behavior. The 
higher-frequency feature quickly decreases its spectral weight with
increasing coupling constant whereas the lower-frequency feature
does the opposite. Besides, the numerically exact calculations of the OC
(DMC) follow the prediction of the extended MFF for $\alpha < 6$,
while they are in fair agreement with SCE for $\alpha >10$. We conclude
that nonadiabaticity destroys the FC classification for $\alpha < 10$
while the FC principle rapidly regains its validity at large coupling
strengths due to the fast growth of the energy separation between
the initial and final states of the optical transitions. Furthermore,
both adiabatic FC and nonadiabatic dynamical excitations coexist in the
intermediate e-ph coupling regime, $6 <\alpha <10 $. The
crossover is controlled by the adjustment time $\tau_{ic} \approx \hbar / 
\mathcal{D}$, set by the typical nonadiabatic energy $\mathcal{D}$ (see
text below).

In the Fr\"{o}hlich polaron model\cite{Frohlich} the electron ($\vec{r}$
and $\vec{p}$ are the position and momentum operators) is scattered by
phonons ($a_{\vec{q}}^{\dagger }$ denotes the creation operator with wave
number $\vec{q}$) with e-ph interaction vertex $M_{q}=i\hbar \omega
_{0}\left( R_{p}4\pi \alpha /q^{2}V\right) ^{1/2}$: 
\begin{equation}
H=p^{2}/2m+\hbar \omega _{0}\sum_{\vec{q}}a_{\vec{q}}^{\dagger }a_{\vec{q}%
}+\sum_{\vec{q}}[M_{q}e^{i\vec{q}\cdot \vec{r}}a_{\vec{q}}+h.c.].
\end{equation}%
Here $\alpha $ is the dimensionless coupling constant, $R_{p}=\left( \hbar
/2m\omega _{0}\right) ^{1/2}$, and $V$ is the volume of the system. The band
mass of the electron $m$, Planck's constant $\hbar $, the
electron charge and the dispersionless longitudinal optical phonon
frequency $\omega _{0}$ are set below to unity. Although the OC of
this model attracted attention for years\cite{Devreese}, there exists no
analytic approach giving a satisfactory description for all coupling
regimes. The most successful approach is that based on the Feynman
path integral technique\cite{dsg} (DSG), where the OC is calculated starting
from the Feynman variational model (FVM)\cite{Feynman} for the polaron and
using the path integral response formalism\cite{dsg1}. Subsequently the
path integral approach was rewritten in terms of the memory function
formalism (MFF)\cite{Devreese1}. These approaches are completely successful
at small e-ph couplings, are able to identify some of the
excitations at intermediate and strong e-ph couplings without 
reproducing the  broad structures present in DMC data \cite{Andrei}.

\textit{Extended memory function formalism}. In order to solve the
aforementioned problem regarding the description of the OC main peak line
width at intermediate e-ph couplings, we modified the DSG approach
to include additional dissipation processes, whose strength is fixed by
an exact sum rule. Within MFF\cite{Mori} the interaction of the charge
carriers with the free phonon oscillations can be expressed in terms of the
electron density-density correlation function, $\chi (\vec{q},t)=-i\theta
(t)\left\langle exp\left[ i\vec{q}\cdot \vec{r}(t)\right] exp\left[ -i\vec{q}%
\cdot \vec{r}(0)\right] \right\rangle $, which is evaluated in a direct
way\cite{Devreese1} using the FVM, where the electron is coupled via a
harmonic force to a fictitious particle that simulates the phonon
degrees of freedom. Within this procedure the electron density-density
correlation function takes the form: $\chi _{m}(\vec{q},t)=-i\theta (t)exp%
\left[ -iq^{2}t/2M\right] exp\left[ -q^{2}R(1-e^{-ivt})/2M\right] $, where $%
R=(M-1)/v$, $M$ (the total mass of electron and fictitious particle), and $v$
are determined variationally within the path integral approach\cite{Feynman}%
. The associated spectral function $A_{m}(\vec{q},\omega )=-2\Im \chi _{m}(%
\vec{q},\omega )$ is a series of $\delta $ functions centered at $q^{2}/2M+nv
$ ($n$ is integer). Here $q^{2}/2M$ represents the energy of the center of
mass of electron and fictitious particle, and $v$ is the energy gap between
the levels of the relative motion. To include dissipation we introduce
here a finite lifetime for the states of the relative motion, which can be
considered as the result of the residual e-ph interaction not included into
the FVM. To this end, in $\chi _{m}(\vec{q},t)$ we replace the factor $%
exp\left[ -ivt\right] $ with $(1+it/\tau )^{-v\tau }$ which leads to the
replacement of $\delta $ functions by Gamma functions with mean value and
variance given respectively by $q^{2}/2M+nv$ and $nv/\tau $. The parameter
of dissipation $\tau $ is not an adjustable parameter but is determined by
the third sum rule for $A(\vec{q},\omega )$, which is additional to the
first two ones that are already satisfied in the DSG model without
damping. As expected, $\tau $ turns out to be of the order of $\omega
_{0}^{-1}$. If broadening of the oscillator levels is neglected, $\tau
\rightarrow \infty $, the DSG results\cite{dsg,Devreese1} are recovered.
Within this latter approach\cite{dsg,Devreese1} the polaron OC was
previously interpreted in terms of relaxed excited states transitions,
FC-transitions and transitions to scattering states \cite{Devreese2,GSD}.

\textit{Strong coupling expansion}. In the limit of strong coupling
strengths the adiabatic lattice deformation is large and the lattice
kinetic energy can be regarded as a perturbation. Within the Landau and
Pekar\cite{Pekar}(LP) approach adiabatic lattice displacements are taken
into account by a unitary transformation and the Hamiltonian is divided
into two contributions, $H=H_0+H_I$: the first one describes the
electron oscillation in a self-consistent quadratic potential with frequency 
$\omega_1^{LP} = 4 \alpha^2 / 9\pi$ and the second one represents the
residual e-ph interaction. To get a quantitative estimate for the
characteristic frequency of the quadratic potential, we improve the LP value
considering the effects of the translational invariance and the residual
interaction $H_I$\cite{noi}. It turns out that the frequency $\omega_1 = (4
\alpha^2 / 9\pi-3.8)$ differs from that of LP $\omega_1^{LP}$ by a
constant shift 3.8. Then, starting from the Kubo formula, taking into
account all multiphonon processes and neglecting recoil as well as
correlation between the emission and absorption of successive phonons
in all orders of the
perturbation $H_I$, one arrives to the following expression for the real
part of the OC 
\begin{equation}
\Re \sigma(\omega)=\sigma_0 \omega \sum^{\infty}_{n=0} \frac {e^ {-\omega_s} 
} {n!} \left( \omega_s \right)^n  \delta \left[\omega-\omega_2 -n\right],
\label{1r}
\end{equation}
resembling that expected for an exactly solvable independent oscillators
model \cite{Mahan}. Parameters, specific for the Fr\"{o}hlich polaron model,
are $\omega_2=\omega_1-\omega_s$, $\sigma_0=\pi /2 \omega_1$ and $%
\omega_s=\alpha \left( \omega_1/ 16 \pi \right)^{1/2}$. The parameter $%
\omega_1$ is the FC transition energy, and $\omega_s$ is the energy
shift due to lattice relaxation. Naturally, for large enough coupling
strengths the envelope of the Poisson distribution (\ref{1r}) is well
described by the Gaussian 
\begin{equation}
\Re \sigma(\omega)= {\omega} \frac {\sigma_0} {(2 \pi \omega_s )^{1/2}} \exp
\left\{-\frac {(\omega-\omega_1)^2} {2 \omega_s } \right\}.  \label{2r}
\end{equation}

%%%%%%%%%%%%%%%%%%%%%%%%%%%%
\begin{figure*}
\centering \includegraphics[width=6.2in]{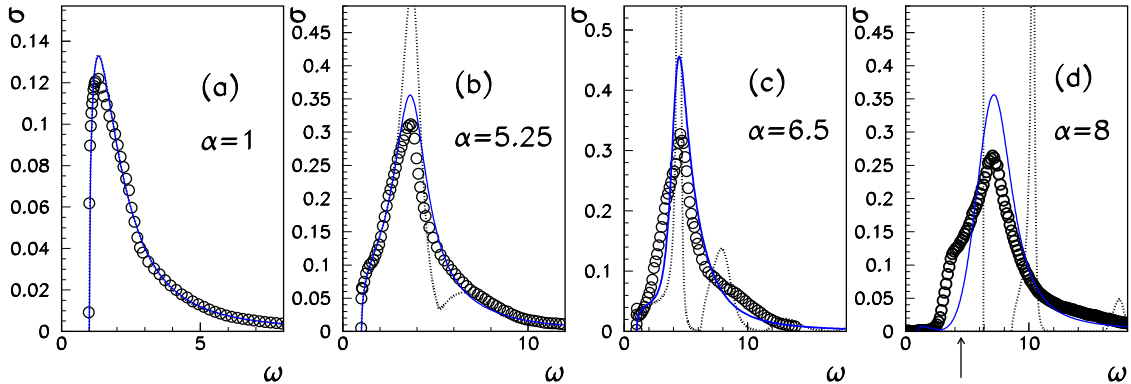}
\caption{Comparison of the optical conductivity calculated within the
DMC method (circles), extended MFF (solid line) and DSG\protect\cite%
{dsg,Devreese1} (dotted line), for four different values of $\protect\alpha$%
. The arrow indicates the lower-frequency feature in the DMC data. }
\label{figa_graph}
\end{figure*}
%%%%%%%%%%%%%%%%%%%%%%%%%%%%

%%%%%%%%%%%%%%%%%%%%%%%%%%%%
\begin{figure*}
\centering \includegraphics[width=6.2in]{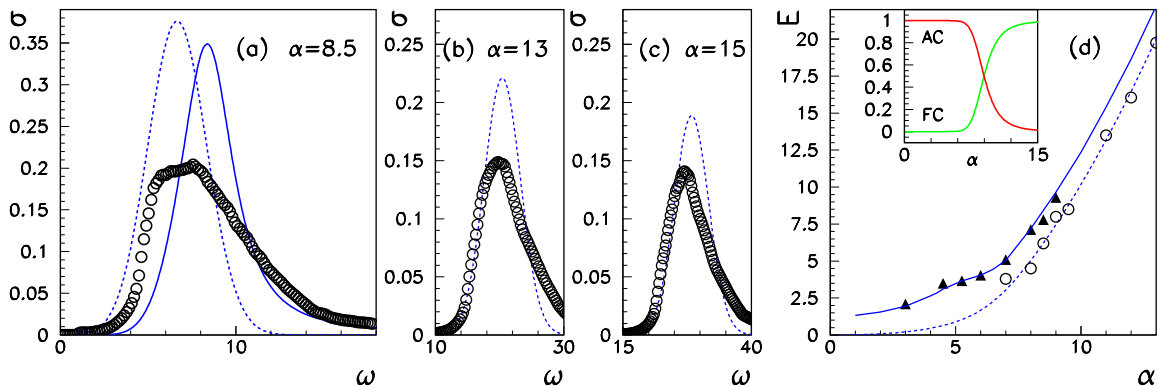}
\caption{a), b) and c) Comparison of the optical conductivity calculated
within the DMC method (circles), the extended MFF(solid line) and
SCE (dashed line) for three different values of $\protect\alpha$. d) The
energy of the lower- and higher-frequency features obtained by DMC
(circles and triangles, respectively) compared with the FC transition
energy as calculated with the SCE (dashed line) and with the energy of
the peak obtained from the extended MFF (solid line). In the inset the
weights of Franck-Condon and adiabatically connected transitions are shown
as a function of $\protect\alpha$. We have used for $\protect\eta$ the value $1.3$. }
\label{figb_graph}
\end{figure*}
%%%%%%%%%%%%%%%%%%%%%%%%%%%%
%[width=.9\columnwidth]

\textit{Interpretation of the DMC results}. As expected, in the weak
coupling regime (Fig.1a), both our extended MFF with phonon broadening
and DSG\cite{dsg} are in very good agreement with the DMC data\cite%
{Andrei}, showing significant improvement with respect to the weak coupling
perturbation approach\cite{Lang} which provides a good description of the
OC spectra only for very small values of $\alpha $\cite{note}. For $4\leq
\alpha \leq 8$, where DSG underestimates the peak width (Figs.~1b to 1d),
the damping, introduced in the extended MFF approach, becomes crucial.
Results of the extended MFF are accurate for the peak energy and quite
satisfactory for the peak width (Fig.~1b-d).

However, we observe a very interesting and unexpected behavior in the
coupling regime $6<\alpha<10$. Two features are present in the OC
given by DMC: the position of the lower-frequency peak (or shoulder)
corresponds to the predictions of the SCE (eqs.\ref{1r},\ref{2r})
 while that for the higher-frequency peak follows the extended MFF
value (Fig.~2a). The higher-frequency feature rapidly decreases its
intensity with increasing $\alpha$ and, at large values of $\alpha$
(Figs.~2b and 2c), the OC given by DMC is in fair agreement with
the SCE results, which are strongly dominated by the FC transitions.
Finally, comparing the peak and shoulder energies, obtained by DMC, with
the peak energies, given by MFF, and the FC transition energies from the
SCE (Fig.~2d), we conclude that as $\alpha$ increases from $6$ to $10$
the spectral weights rapidly switch from the dynamic regime, where the
lattice follows the electron motion, to the adiabatic regime dominated by FC
transitions, where the nuclei are frozen in their initial
configuration.

\textit{Breakdown of the FC picture}. In order to support this scenario we
present an analytical estimation of the FC breakdown based on the following
arguments. The approximate adiabatic states $\chi_{i,\beta}(Q)\psi_{i}(%
\mathbf{r},Q)$, where $i$ is the electronic index and $\chi_{i,\beta}(Q)$
is the eigenfunction of the lattice connected with the electron wave
function $\psi_{i}(\mathbf{r},Q)$, are not exact eigenstates of the
system. These states are mixed by nondiagonal matrix elements of the
nonadiabatic operator $\mathcal{D}$ and the exact eigenstates are linear
combinations of the adiabatic wave functions. Being interested in the
properties of transitions from the ground ($G$) state to an excited (EX)
 state, whose energy corresponds to that of the OC peak, we consider mixing
of only these states and express the exact wave functions $\Psi_{G,EX}(%
\mathbf{r},Q)$ as linear combinations\cite{Brovman,KiMi93} $\Psi_{G,EX} =
\xi_{g,\beta}^{G,EX} \chi_{g,\beta} \psi_{g} +
\xi_{ex,\beta^{\prime}}^{G,EX} \chi_{ex,\beta^{\prime}} \psi_{ex}$ of the
adiabatic groundstate ($\chi_{g,\beta} \psi_{g}$) and the adiabatic
excited state ($\chi_{ex,\beta^{\prime}} \psi_{ex}$). The superposition
coefficients are determined from standard techniques\cite{Brovman,KiMi93}
where the nondiagonal matrix elements of the nonadiabatic operator\cite%
{Brovman} are expressed in terms of matrix elements of the kinetic energy
operator $M$, the energy gap between the excited and ground states $\Delta E
= E_{ex}-E_{g} $, and the number $n_{\beta}$ of phonons in the adiabatic
state: 
\begin{equation}
\mathcal{D^{ \pm} }= M (\Delta E)^{-1} \sqrt{n_{\beta} + 1/2 \pm 1/2} +
M^{2} (\Delta E)^{-2}.
\end{equation}
The extent to which the lattice can follow a transition between
electronic states, depends on the degree of mixing between initial and final
exact eigenstates through the nonadiabatic interaction. If the states are
strongly mixed, the adiabatic classification has no sense, the FC
transitions have no place and the lattice is adjusted to the change of the
electronic states during the transition. In the opposite limit the
adiabatic approximation is valid and FC processes dominate. An estimate
for the weight of the FC component is 
\begin{equation}
I_{FC} = 1 - 4 \left\vert \xi_{g,\beta}^{EX} \xi_{ex,\beta^{\prime}}^{G}
\right\vert^2 \; ,
\end{equation}
which is equal to unity in case of zero mixing and zero in case of maximal
mixing. The weight of the adiabatically connected (AC) transition $%
I_{AC}=1-I_{FC}$ is defined accordingly. The non diagonal matrix element 
$M$ is proportional to the square root of $\alpha$ with a coefficient $\eta$
of the order of unity. In the strong coupling regime, assuming that $\Delta
E=\omega_1$, and $n_{\beta} \approx \Delta E$ ($n_{\beta} \gg 1$), one gets 
\begin{equation}
I_{FC} = \left[ 1 + 4 (\tau_{mp}/\tau_{ic})^{2} \right]^{-1} \; ,
\end{equation}
where $\tau_{mp}= 1/\Delta E $ and $\tau_{ic}= 1/D$. For $\eta$ of the order
of $1$, one obtains a robust qualitative description of a rather fast
switch from AC- to FC-dominated  transitions, when $I_{FC}$ and $I_{AC}$
exchange half of their weights in the range of $\alpha$ from 7 to 10 (see
inset of Fig.~2d). The physical reason for such a quick change is the
faster growth of the energy separation $\Delta E \sim \alpha^2$ compared
to that of the matrix element $M \sim \alpha^{1/2}$. This switch has
nothing to do with the self-trapping phenomenon where crossing and
hybridization of the ground state and an excited state occurs. The AC-FC
switch is a property of transitions between different states and is
related to the choice whether the lattice can or cannot follow
adiabatically the change of electronic state at the transition.

\textit{Conclusions}. Comparing numerically exact data on the optical
conductivity, obtained by the Diagrammatic Monte Carlo method, with results
from our Extended Memory Function approach with phonon broadened levels,
which takes dynamic lattice relaxation into account, and results of the
strong coupling expansion we found that the Franck-Condon picture breaks
down at $\alpha<10$. The breakdown of the Franck-Condon picture is caused by
nonadiabatic mixing of initial and final states, which destroys the
Franck-Condon classification scheme and, hence, the excitation processes
with dynamic adjustment of the lattice start to dominate. Finally we
find evidence for an intermediate coupling regime $6<\alpha<10$ where static
and dynamic lattice responses coexist.

\end{document}